# Scientific Hypothesis Generation by Large Language Models: Laboratory Validation in Breast Cancer Treatment

Abbi Abdel-Rehim[1], Hector Zenil[2,3], Oghenejokpeme Orhobor[4], Marie Fisher[5], Ross J. Collins[5], Elizabeth Bourne[5], Gareth W. Fearnley[5], Emma Tate[5], Holly X. Smith[5], Larisa N. Soldatova[6], Ross D. King*[,1,7]

[1] Department of Chemical Engineering and Biotechnology, University of Cambridge, CB3 0AS, U.K.

[2] Research Departments of Biomedical Computing and Digital Twins, School of Biomedical Engineering and Imaging Sciences, King's Institute for AI, King's College London, SE1 7EU, U.K.

[3] The Alan Turing Institute, British Library, London, NW1 2DB, U.K.

[4] The National Institute of Agricultural Botany, Cambridge, CB3 0LE, U.K.

[5] Arctoris Ltd, Oxford, OX14 4SA, UK.

[6] Department of Computing, Goldsmiths, University of London, SE14 6NW, U.K.

[7] Department of Computer Science and Engineering, Chalmers University, S-412 96 Göteborg, Sweden.

*Ross King. Email: rk663@cam.ac.uk

## ABSTRACT

Large language models (LLMs) have transformed AI and achieved breakthrough performance on a wide range of tasks. In science the most interesting application of LLMs is for hypothesis formation. A feature of LLMs, which results from their probabilistic structure, is that the output text is not necessarily a valid inference from the training text. These are termed "hallucinations" and are harmful in many applications. In science some hallucinations may be useful: novel hypotheses whose validity may be tested by laboratory experiments. Here, we experimentally test the application of LLMs as a source of scientific hypotheses using the domain of breast cancer treatment. We applied the LLM GPT4 to hypothesize novel synergistic pairs of FDA-approved non-cancer drugs that target the MCF7 breast cancer cell line relative to the non-tumorigenic breast cell line MCF10A. In the first round of laboratory experiments GPT4 succeeded in discovering three drug combinations (out of twelve tested) with synergy scores above the positive controls. GPT4 then generated new combinations based on its initial results, this generated three more combinations with positive synergy scores (out of four tested). We conclude that LLMs are a valuable source of scientific hypotheses.

## INTRODUCTION

The world has been stunned by the success of Large Language Models (LLMs). They have achieved breakthrough performance on a wide range of conversation-based tasks that previously required human intelligence. The overall architecture of LLMs is remarkably simple: they map input token strings to output token strings using deep neural networks (DNNs). Their power comes from being trained on very large general corpuses (substantial percentages of the whole text-based internet), and the use of very large numbers of both tokens (> 10^4) and parameters (> 10^12). The success of LLMs is surprising given that they don't use any explicit model of the world, nor explicit internal symbols, nor do they have any physical grounding in the world. All of these were assumed by most AI scientists to be essential for such intelligent behaviour.

LLMs can be applied to many aspects of science: to summarize texts (*1, 2*), to analyze data (*3*), to write papers and code (*4*), to formalize knowledge (*5*), to answer questions (*6*), etc. However, the most exciting application for LLMs in science is for generating novel hypotheses. Despite the clear potential of LLMs for hypothesis generation their utility for hypothesis generation has been little investigated.

The architecture of LLMs entails that the output string is the most likely one given the input string and the training data. For science these strings may be interpreted as scientific hypotheses. Due to their internal complexity and sophistication LLMs have the potential to go beyond existing text-based scientific hypothesis generation tools (*7, 8, 9*). The generation of hypotheses by LLMs is closely related to the phenomena of "hallucinations". These are LLM outputs that are not valid inferences from the training data. Some hallucinations may be simply factually wrong. For others, their validity is uncertain. Hallucinations are a serious problem in many applications (*9*). For example, in science it is not acceptable to hallucinate (make up) false references. However, in scientific hypothesis generation hallucinations may be useful: probable novel hypotheses whose validity may be objectively tested by laboratory experiments (*10*).

Here we use laboratory experiments to test the utility of the general purpose LLM GPT4 for scientific hypothesis generation (Fig. 1). We employed breast cancer as the test domain. In our experiments breast cancer cells were exemplified by MCF7 (an epithelial breast cancer cell line); non-tumorigenic breast cells were exemplified by the epithelial cell line MCF10A. We provided GPT4 with a prompt that had several aims: 1) Identify novel drug combinations that would have a significant impact on MCF7 cell lines; 2) Avoid harming MCF10A the control cell line; and 3) Design combinations that were possibly synergistic. We also had additional requirements related to the drugs themselves: at least one of the drugs in every pair should not be an anti-neoplastic drug, and that the drugs should be affordable, accessible, and preferably FDA-approved. Prompts and the list of complete hypotheses are found in supplementary materials (Fig. S1-S3, Table S1). Interestingly, all the drug combinations hypothesized were exclusively non-cancer drugs (suggesting a possible limitation in GPT4 understanding of its instructions). The combinations were however novel, and we could not find any of them reported for breast cancer in the literature. We did find that several of the drugs had been investigated in the cancer literature, and that several of the drugs had been tested against MCF7 (Supplementary materials Appendix).

In addition to hypothesizing drug combination, we prompted GPT4 to provide two positive controls that are commonly used against breast cancer in clinic and likely have an impact on MCF7; as well as two negative control combinations that would be unlikely to cause harm to MCF7 (Fig. S2, S3). It may have been wiser to select the controls ourselves. But we judged that GPT4 did a fair job in its selections (Table 1b).

## RESULTS

Using the method described below we screened the twelve pairs of compounds proposed by GPT4 (Table 1a). We investigated two properties of the pairs: 1) The specificity of the combination for MCF7 v MCF10A, and 2) The additivity/synergy of the combination. (Additivity occurs when the combination of the effects of two drugs is not less than either of the two drugs acting independently. Synergy describes the situation when the effect of the combination is greater than that of the most effective drug (highest single agent)). To determine drug additivities/synergies we employed SynergyFinder 3.0 to calculate HSA (highest single agent) synergy scores for all combinations (Table 2). There were six additive interactions combinations with positive synergy scores for MCF7: itraconazole + atenolol, simvastatin + disulfiram, dipyridamole + mebendazole, furosemide + mebendazole, disulfiram + hydroxychloroquine, and the positive control doxorubicin + cyclophosphamide. The initial three hypothesised combinations resulted in HSA scores surpassing those of the positive controls. Synergistic areas were found within the drug response matrices belonging to ten out of twelve of the hypothesized drug combinations (Table S2). We found that eight out of the twelve hypothesised combinations resulted in a higher HSA score in varying degrees for MCF7 compared to MCF10A (Table 2, cf. Table 2 and Table S3). In Table S12 we summarize the literature on the hypotheses proposed by GPT4 and the anti-cancer properties of the drugs selected. We found underlying support in literature for three out of the latter screened six combinations with positive synergy scores, while the remaining three remain unclear.

To better understand the utility of the paired compounds we tested the individual drugs (Table 3). From the drugs in the positive controls pairs only doxorubicin was found to result in an IC50 value below the maximum dose of 25uM in both cell lines. For MCF7 there were five additional drugs that resulted in IC50 values below the same threshold, with Disulfiram and Niclosamide showing comparatively high toxicity (Table 3). Several more drugs were toxic to the cell lines, but failed to reduce the viability to such an extent where an IC50 value could be derived (Table S4 and S5). In total, twelve out of the eighteen non-control drugs showed toxicity towards MCF7: celecoxib, chloroquine, dipyridamole, disulfiram, hydroxychloroquine, itraconalzole, mebendazole, niclosamide, quinacrine, sildenafil and simvastatin. Out of these drugs, dipyridamole, disulfiram, mebendazole and quinacrine showed high specificity towards MCF7 (cf. Table S4 and S5). While many of these drugs had been studied in cancer cell lines, they are not cancer drugs. Fulvestrant, a positive control cancer drug also showed preference for MCF7.

Eleven out of eighteen compounds reduced viability of the control cell line MCF10A. When excluding the highest concentration of 25 uM, these numbers change to 6/18 and 8/18 compounds for MCF7 and MCF10A respectively. The ten drugs that showed highest toxicity towards MCF7 were re-screened to achieve sufficient replicates (n=3) in order to validate their toxicity.

In an additional experiment, twelve drugs were retested from the first round (n >=3 replicates). An ANOVA two-way test with three replicates was used to calculate significance of changes to viability compared to the internal control drug Allopurinol for both cell lines (S6, S7). Two of the retested drugs were initially used as positive controls for MCF7 (doxorubicin and fulvestrant). Out of twelve retested drugs, dipyridamole, disulfiram, niclosamide and quinacrine significantly reduced the viability of MCF7 when considering concentrations up to 3.84 µM. The two positive control drugs doxorubicin and fulvestrant also targeted MCF7. When considering all concentrations (including 25 µM which is quite high), all but hydroxychloroquine results in a significant impact on MCF7. Despite this, the toxicity of hydroxychloroquine at 25 uM is persistent and substantial. IC50 values could only be calculated for 5/11 compounds (Table S8), disulfiram and doxorubicin showing sub micromolar IC50 values of 0.059 and 0.3 µM respectively. This was followed by niclosamide at 1.22 µM, quinacrine at 4.71 µM and chloroquine at 10.62 µM. However, the remaining compounds were still toxic to MCF7 (Table S9). Calculated IC50, and viability values for MCF10A can be found in Tables S8 and S10.

<table>
<tr><th></th><th>Drug1</th><th>Drug2</th></tr>
<tr><td rowspan="2">1</td><td>Doxorubicin</td><td>Cyclophosphamide</td></tr>
<tr><td colspan="2">Positive Control.<br><br>Combination FDA approved to treat Breast Cancer<br><br>Doxorubicin is an anthracycline that intercalates into DNA and inhibits topoisomerase II, causing DNA damage. Cyclophosphamide is an alkylating agent that causes DNA damage.<br><br>"The combination targets DNA integrity through multiple mechanisms, which may be effective in MCF7 cells with high proliferative capacity."</td></tr>
<tr><td rowspan="2">2</td><td>Fulvestrant</td><td>Palbociclib</td></tr>
<tr><td colspan="2">Positive Control.<br><br>Individually FDA approved to treat Breast Cancer<br><br>Fulvestrant is a selective estrogen receptor degrader (SERD) that blocks and degrades estrogen receptors. Palbociclib is a CDK4/6 inhibitor that blocks cell cycle progression.<br><br>"The combination targets both estrogen signaling and cell cycle progression, which may be effective in estrogen receptor-positive MCF7 cells."</td></tr>
<tr><td rowspan="2">3</td><td>Allopurinol<br>(Xanthine Oxidase Inhibitor)</td><td>Omeprazole<br>(Proton Pump Inhibitor)</td></tr>
<tr><td colspan="2">Negative Control<br><br>"Allopurinol is used to treat gout and hyperuricemia, and omeprazole is used to reduce stomach acid. Neither drug targets pathways relevant to MCF7 breast cancer cell growth or survival, and they are not expected to have an effect on MCF7 cells."</td></tr>
<tr><td rowspan="2">4</td><td>Diphenhydramine (Antihistamine)</td><td>Omeprazole (Proton Pump Inhibitor)</td></tr>
<tr><td colspan="2">Negative Control<br><br>"Diphenhydramine and cetirizine are antihistamines used to treat allergy symptoms. Neither drug targets pathways relevant to MCF7 breast cancer cell growth or survival, and they are not expected to have an effect on MCF7 cells."</td></tr>
</table>

Table 1b. GPT4 generated positive and negative control combinations.

| Drug 1 | Drug 2 | HSA score (MCF7) | Specificity (MCF7) |
|---|---|---|---|
| Itraconazole | Atenolol | 4.83 | 7.03 |
| Simvastatin | Disulfiram | 3.29 | 1.85 |
| Dipyridamole | Mebendazole | 2.49 | 3.69 |
| Doxorubicin* | Cyclophoshamide* | 1.02 | 3.27 |
| Furosemide | Mebendazole | 0.72 | 6.14 |
| Disulfiram | Hydroxychloroquine | 0.60 | 3.51 |
| Acarbose | Itraconazole | -1.36 | -1.33 |
| Disulfiram | Sildenafil | -1.63 | 0.85 |
| Allopurinol | Chloroquine | -1.87 | 2.24 |
| Celecoxib | Quinacrine | -2.21 | -3.27 |
| Fulvestrant* | Palbociclib* | -2.59 | -0.49 |
| Memantine | Niclosamide | -2.61 | -2.23 |
| Disulfiram | Cimetidine | -3.06 | -8.17 |
| Allopurinol** | Omeprazole** | -3.85 | -6.2 |
| Atrovastatin | Metronidazole | -4.84 | -6.3 |
| Diphenhydramine** | Cetirizine** | -9.25 | -6.28 |

Table 2. HSA synergy score for each combination. “Specificity” indicates synergy score differences between the cell lines ($HSA_{MCF7}$-$HSA_{MCF10A}$). *Positive controls, **negative controls. Combinations selected for further validation are marked in bold. The combinations in blue have positive synergy scores .

| Drug | MCF7 IC50 (uM) | MCF7 p-val | MCF10A IC50 (uM) | MCF10A p-val |
|---|---|---|---|---|
| Allopurinol** | >25 | ---------- | >25 | ---------- |
| Atenolol | >25 | **0.003** | >25 | 0.118 |
| Celecoxib | 5.325 | **0.046** | 22.573 | 0.185 |
| Disulfiram | 0.204 | **0.008** | >25 | 0.095 |
| Fulvestrant* | >25 | **0.020** | >25 | 0.430 |
| Itraconazole | >25 | **0.021** | >25 | 0.077 |
| Sildenafil | >25 | **0.011** | >25 | 0.212 |
| Cimetidine | >25 | **0.012** | >25 | **0.023** |
| Mebendazole | >25 | **0.025** | 15 | **0.018** |
| Metronidazole | >25 | **0.039** | >25 | **0.031** |
| Atorvastatin | >25 | 0.131 | 3.795 | **0.009** |
| Chloroquine | >25 | 0.202 | >25 | **0.030** |
| Doxorubicin* | 0.303 | 0.054 | 0.435 | **0.034** |
| Memantine | >25 | 0.834 | >25 | **0.022** |
| Niclosamide | 0.699 | 0.066 | 0.061 | **0.021** |
| Acarbose | >25 | 0.251 | >25 | **0.019** |
| Cetirizine** | >25 | 0.210 | >25 | 0.257 |
| Cyclophosphamide* | >25 | 0.276 | >25 | 0.499 |
| Diphenhydramine** | >25 | 0.684 | >25 | 0.500 |
| Dipyridamole | >25 | 0.056 | >25 | 0.093 |
| Furosemide | >25 | 0.246 | >25 | 0.188 |
| Hydroxychloroquine | >25 | 0.118 | >25 | 0.944 |
| Omeprazole** | >25 | 0.082 | >25 | 0.245 |
| Palbociclib* | >25 | 0.414 | >25 | 0.650 |
| Quinacrine | 3.848 | 0.082 | 10.183 | 0.116 |
| Simvastatin | 5.634 | 0.106 | 7.17 | 0.120 |

Table 3. Single drug treatments. For either cell line, the IC50 values were calculated and declared (n=1). Significance (p-value) of changes evoked by drug treatments when compared to positive control (Allopurinol) are also reported. ANOVA two-way significance is used for the samples, most of which are single replicates. P-values are declared for both cell lines. *=Positive controls, ** = negative controls. Numbers in bold significant <0.05. Blue significant against MCF7. Red significant against MCF10A. Green significant against MCF7 and MCF10A

After the results from the first round of experiments were complete we investigated whether GPT4 could improve its hypotheses through use on the results from its initial hypotheses. We provided GPT4 with a summary of the results from the primary screen (Fig. S4), and prompted GPT4 to consider combinations containing drugs from the positive controls as well. GPT4 hypothesized four combinations based on this information: disulfiram + fulvestrant, disulfiram + mebendazole, mebendazole + quinacrine, and disulfiram + quinacrine (Table 4). In addition we re-tested three combinations that resulted in positive synergy scores from the primary screening achieving more robust results, these combinations were disulfiram + simvastatin, disulfiram + hydroxychloroquine, and dipyridamole + mebendazole. Out of the seven combinations screened in the second iteration, six combinations showed varying degrees of synergy within the response matrices (Table 5). Of the newly hypothesized pairs we found three pairs with positive synergy scores: mebendazole + quinacrine, disulfiram + fulvestrant, and disulfiram + quinacrine. The remaining three re-tested combinations also showed consistent positive scores. The three combinations with the highest HSA scores also showed specificity (>1 HSA score) towards MCF7. It is worth mentioning that the most synergistic 3x3 dose response window resulted in one of the combinations (disulfiram + simvastatin) having a synergy score >10. Plots showing HSA synergy graphs for MCF7 and MCF10A derived from Synergyfinder 3.0 can be found in Supplementary Materials Appendix C1 and C2. HSA scores for MCF10A can be found in Table S11. When comparing the most synergistic dose-response windows across both cell lines, there were three combinations that showed substantially higher synergy against MCF7: Quinacrine + mebendazole (ΔHSA=2.73), mebendazole+dipyridamole (ΔHSA=3.99) and simvastatin+disulfiram (4.01) (Appendix D). In addition, the two latter combinations also showed areas of selective toxicity towards the MCF7 cell line.

| Drug 1 | Drug 2 | HSA score | HSA score (max) | Specificity |
|---|---|---|---|---|
| Disulfiram | Simvastatin | 4.75 | 10.58 | 2.41 |
| Disulfiram | Fulvestrant | 1.81 | 4.60 | 0.03 |
| Disulfiram | Quinacrine | 1.53 | 4.47 | 0.6 |
| Dipyridamole | Mebendazole | 1.10 | 5.26 | 3.60 |
| Disulfiram | Hydroxychloroquine | 1.08 | 3.06 | -1.86 |
| Mebendazole | Quinacrine | 0.56 | 5.54 | 1.09 |
| Disulfiram | Mebendazole | -2.49 | 0.13 | -1.83 |

Table 5. Calculated HSA synergy scores for each combination. Two HSA scores are provided, the first considers the entire dose-response matrix, while the second (max) is based on the most synergistic 3-by-3 dose-window in the dose-response matrix. Specificity denotes the difference in the overall HSA synergy score between the two cell lines ($HSA_{MCF7}$-$HSA_{MCF10A}$) where positive values indicate higher synergy scores for MCF7.

A final query was made to GPT4 requesting future experiments based on the final results. Three drug combinations were recommended: disulfiram + itraconazole, mebendazole + cimetidine, and quinacrine + celecoxib. Hypotheses for these combinations are reported in Table S13. Disulfiram + itraconazole were hypothesized to synergise based on increased oxidative stress and the inhibition of the hedgehog pathway. Mebendazole and cimetidine were also hypothesized to synergise due to their targets being involved in cell cycle progression and growth. The final combination quinacrine + celecoxib had been tested in the initial experiment, suggesting that GPT4 had already “forgotten” its previous recommendations.

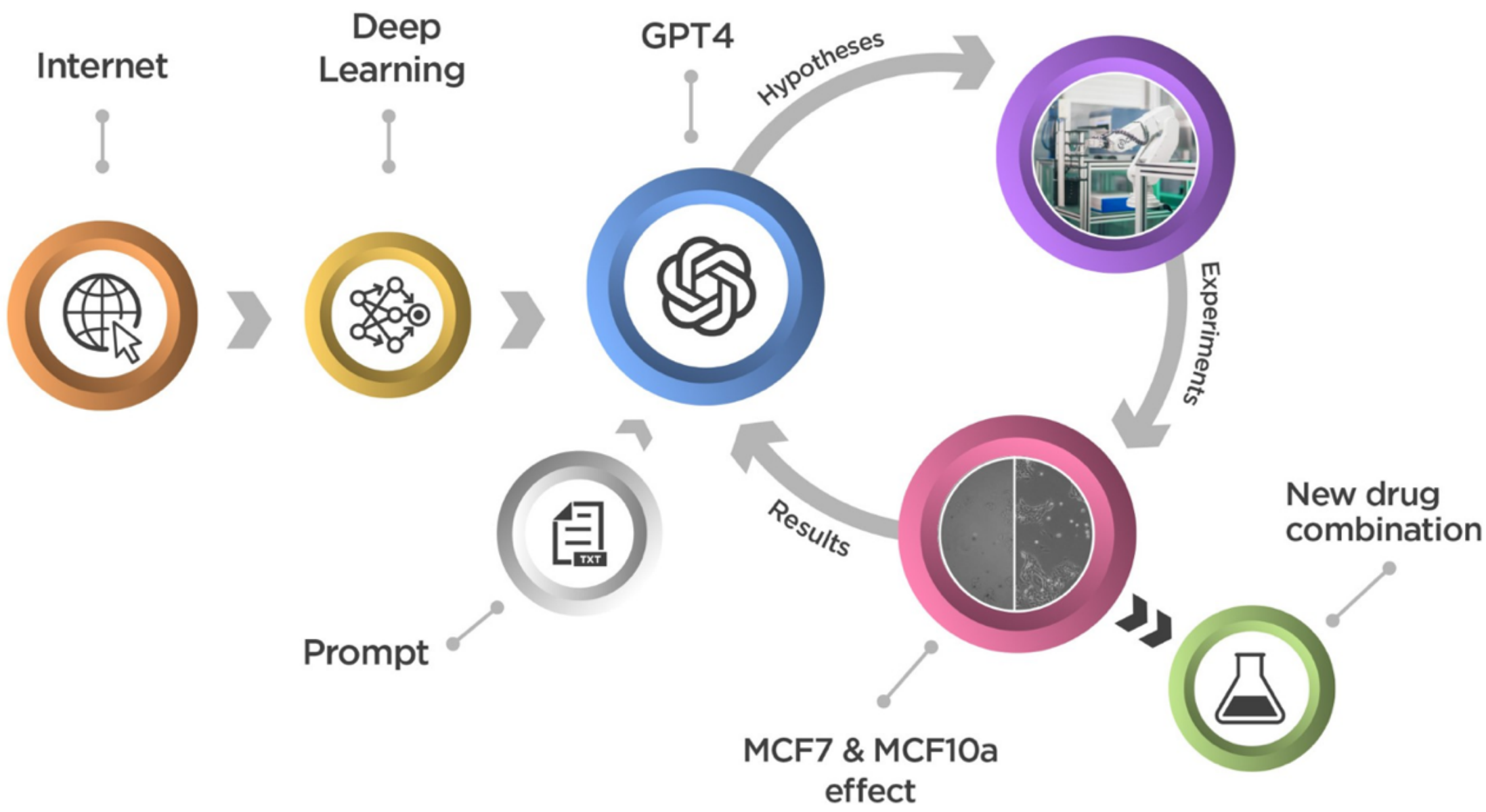

Figure 1. The overall structure of our experiments. GPT4 was previously trained on data on a large fraction of the text on the internet. We engineered prompts to request from GPT4 synergistic pairs of drugs that are toxic to the breast cancer cell line MCF7, but not to the non-cancer breast cell line MCF10a. These are the hypotheses that we experimentally tested using laboratory automation.

## DISCUSSION

It is unclear to what extent GPT4 "understood" its prompt for hypothesis formation. This epistemological uncertainty is show in the relationship between the explanation of why a pair of drugs would target MCF7 rather than MCF10A (Table S1), and the explanation why MCF10A would not be targeted, where the MCF10A hypotheses are simply negations of the MCF7 ones. More convincing explanations for not targeting MCF10A would have provided us with more confidence in GPT4's understanding, and the utility of its hypotheses. The hypotheses also exhibited "hallucinations" in their explanations. This is most clearly illustrated by GPT4's hypothesis that itraconazole would "disrupt(ing) cell membrane integrity". This explanation presumably originated from the fact that itraconazole inhibits ergosterol synthesis, which disrupts cell membrane integrity. The factual error is that ergosterol synthesis is not present in mammalian cells. We asked GPT4 "is ergosterol synthesis present in mammalian cells". It replied "No, ergosterol synthesis is not present in mammalian cells. Ergosterol is a sterol found in the cell membranes of fungi and some protozoa, playing a role similar to cholesterol in mammalian cells..."

This study focused on GPT-4, but there are several other LLMs available. We compared the outputs from GPT-4, Gemini, and the specialised LLM PubMedGPT. The results revealed both similarities and notable differences in the selected drugs and their subsequent combinations (Appendix D). It was evident that LLMs generated non-uniform distributions in their drug suggestions, with certain drugs being consistently selected across models, while pairs exhibit greater variability. Despite this diversity, the suggestions remain consistent in their underlying choices. The specialised LLM PubMedGPT recommended different drug pairs but with a strong overlap in core drug selections—an aspect that has both advantages and drawbacks. In future studies, it may be useful to analyse pair frequency distributions per LLM.

We selected cancer treatment as our test domain because every cancer patient ideally requires a scientific research project to understand how best to treat them. In the past this was prohibitively expensive and out of reach for normal cancer patients. The cost of science currently has two main components: the cost of the human scientist's intelligence, and the laboratory costs. Now, thanks to the AI revolution, the cost of (machine) scientific intelligence is dropping. Economies of scale could also drive down the cost of the robotics required to automate personalized cancer research. We therefore envision a future where scientific research is cheap enough that every cancer patient can afford treatment based on a personalized research project.

Our empirical results demonstrate that the GPT4 succeeded in its primary task of forming novel and useful hypothesis. We therefore conclude that LLMs are an important source of novel scientific hypotheses for both human scientists and to the increasingly sophisticated AI systems designed to automate science (*11*).

## Acknowledgements

Funding: This work has been supported by the UK Engineering and Physical Sciences Research Council (EPSRC) [EP/R022925/2, EP/W004801/1 and EP/X032418/1], and by the Wallenberg AI, Autonomous Systems and Software Program (WASP) funded by the Alice Wallenberg Foundation.

## Data availability

All experimental data are available in the main text or within the supplementary information.

## Appendix

| | Drug1 | Drug2 |
|---|---|---|
| 1 | Disulfiram (Alcoholism) | Simvastatin (Hypercholesterolemia) |
| | "Disruption of lipid rafts by simvastatin may enhance disulfiram-induced oxidative stress, leading to apoptosis in MCF7 cells." | |
| 2 | Celecoxib (Pain/Inflammation) | Quinacrine (Malaria/Anti-inflammatory) |
| | "Reduced inflammatory signaling by celecoxib may enhance quinacrine-induced impairment of autophagy, leading to apoptosis in MCF7 cells." | |
| 3 | Acarbose (Diabetes) | Itraconazole (Fungal Infections) |
| | "Acarbose-induced glucose deprivation may enhance the effect of itraconazole on disrupting cell membrane integrity, leading to apoptosis in MCF7 cells." | |
| 4 | Dipyridamole (Blood Thinner) | Mebendazole (Parasitic Infections) |
| | "Dipyridamole-induced increase in cAMP levels may enhance the effect of mebendazole on cell cycle arrest, leading to apoptosis in MCF7 cells." | |
| 5 | Atorvastatin (Hypercholesterolemia) | Metronidazole (Bacterial Infections) |
| | "Atorvastatin-induced disruption of lipid rafts may enhance the effect of metronidazole-induced DNA damage, leading to apoptosis in MCF7 cells." | |
| 6 | Allopurinol (Gout) | Chloroquine (Malaria) |
| | "Allopurinol-induced reduction of uric acid may enhance the effect of chloroquine-induced impairment of autophagy, leading to apoptosis in MCF7." | |
| 7 | Cimetidine (Gastric Acid Reducer) | Disulfiram (Alcoholism) |
| | "Cimetidine-induced reduction of gastric acid may enhance the effect of disulfiram-induced oxidative stress and DNA damage in MCF7 cells." | |
| 8 | Memantine (Alzheimer's) | Niclosamide (Parasitic Infections) |
| | "Memantine-induced reduction of glutamate excitotoxicity may enhance the effect of niclosamide-induced energy depletion, leading to cell death in MCF7 cells." | |
| 9 | Furosemide (Diuretic) | Mebendazole (Parasitic Infections) |
| | "Furosemide-induced reduction of sodium reabsorption may enhance the effect of mebendazole on cell cycle arrest, leading to apoptosis in MCF7 cells." | |
| 10 | Sildenafil (Erectile Dysfunction) | Disulfiram (Alcoholism) |
| | "Sildenafil-induced increase in cGMP levels may enhance the effect of disulfiram-induced oxidative stress and DNA damage in MCF7 cells." | |
| 11 | Itraconazole (Fungal Infections) | Atenolol (Hypertension) |
| | "Atenolol-induced reduction of stress signaling may enhance the effect of itraconazole on disrupting cell membrane integrity, leading to apoptosis in MCF7 cells." | |
| 12 | Hydroxychloroquine (Malaria, Rheumatoid Arthritis) | Disulfiram (Alcoholism) |
| | "Hydroxychloroquine-induced impairment of autophagy may enhance the effect of disulfiram-induced oxidative stress and DNA damage in MCF7 cells." | |

Table 1a. GPT4 generated drug combination hypotheses.

<table>
<tr><th></th><th>Drug1</th><th>Drug2</th><th></th></tr>
<tr><td rowspan="2">1</td><td>Disulfiram<br>(Alcoholism)</td><td>Fulvestrant<br>(Breast cancer)</td><td rowspan="2">+</td></tr>
<tr><td colspan="2">"Disulfiram exhibited the highest toxicity to MCF7 cells, and both Disulfiram and Fulvestrant preferentially targeted MCF7 over MCF10A. MCF7 is an estrogen-receptor positive cell line, and Fulvestrant is an estrogen receptor antagonist, which degrades the receptor. This combination could target the estrogen-dependent growth of MCF7 cells. Disulfiram's aldehyde dehydrogenase inhibitory action can further increase its toxicity to MCF7 cells."</td></tr>
<tr><td rowspan="2">2</td><td>Mebendazole<br>(Parasitic Infections)</td><td>Quinacrine<br>(Malaria/Anti-inflammatory)</td><td rowspan="2">+</td></tr>
<tr><td colspan="2">"Both Mebendazole and Quinacrine preferentially targeted MCF7 cells. Mebendazole has anti-cancer properties, potentially by disrupting microtubule function. Quinacrine, though primarily an antimalarial, can exert anti-cancer effects, possibly due to its ability to intercalate with DNA and/or inhibit topoisomerase. This combination might have a multi-pronged effect on MCF7 cells, disrupting cellular division and DNA processes."</td></tr>
<tr><td rowspan="2">3</td><td>Disulfiram<br>(Alcoholism)</td><td>Mebendazole<br>(Parasitic Infections)</td><td rowspan="2">-</td></tr>
<tr><td colspan="2">"As Disulfiram demonstrated the highest toxicity to MCF7 cells and Mebendazole also preferentially targeted MCF7, their combination might exert a compounded effect. The combined microtubule disruption by Mebendazole and the aldehyde dehydrogenase inhibition by Disulfiram could cripple the MCF7 cells at multiple fronts."</td></tr>
<tr><td rowspan="2">4</td><td>Disulfiram<br>(Alcoholism)</td><td>Quinacrine<br>(Malaria/Anti-inflammatory)</td><td rowspan="2">+</td></tr>
<tr><td colspan="2">“Disulfiram exhibited the highest toxicity toward MCF7 cells and has been shown to induce oxidative stress and inhibit proteasomal and NF-κB pathways, which are vital for cancer cell survival. Quinacrine, an antimalarial drug, also possesses anti-cancer properties, including the ability to inhibit NF-κB and autophagy, a survival mechanism often upregulated in cancer cells. The combination is expected to potentiate oxidative stress and inhibit survival pathways more effectively in MCF7 cells. Given that both drugs preferentially targeted MCF7 over MCF10A, their combination might enhance selectivity, exploiting their shared mechanisms for a synergistic effect.”</td></tr>
</table>

Table 4. GPT4 drug combination recommendations based on results from the first screen. Synergy hypotheses are provided for each of the four combinations.